\documentclass{PoS}

\title{Modelling energy-dependent pulsar light curves due to curvature radiation}

\ShortTitle{Energy-dependent pulsar light curve modelling}

\author{\speaker{Monica Barnard}\thanks{A special thank you to the co-authors for their contributions and valuable insights.} \\
        Centre for Space Research, North-West University, Potchefstroom Campus, 
        Private Bag X6001, Potchefstroom 2520, South Africa \\
		E-mail: \email{monicabarnard77@gmail.com}}

\author{Christo Venter \\
		Centre for Space Research, North-West University, Potchefstroom Campus, 
		Private Bag X6001, Potchefstroom 2520, South Africa \\
        E-mail: \email{Christo.Venter@nwu.ac.za}}

\author{Alice K. Harding \\
		Astrophysics Science Division, NASA Goddard Space Flight Center, Greenbelt, 
		MD 20771, USA \\
        E-mail: \email{ahardingx@yahoo.com}}

\author{Constantinos Kalapotharakos \\
		Universities Space Research Association (USRA), Columbia, MD 21046, USA \\		
		Astrophysics Science Division, NASA Goddard Space Flight Center, Greenbelt, 
		MD 20771, USA \\
        University of Maryland, College Park (UMDCP/CRESST), College Park, MD 20742, 
        USA \\
        E-mail: \email{constantinos.kalapotharakos@nasa.gov}}

\abstract{Pulsars emit pulsed emission across the entire electromagnetic spectrum and their light curve phenomenology is strongly dependent on energy. This is also true for the $\gamma$-ray waveband. Continued detections by \textit{Fermi} Large Area Telescope in the GeV band and ground-based Cherenkov telescopes in the TeV band (e.g., Crab and Vela above 1~TeV) raise important questions about our understanding of the electrodynamics and local environment of pulsar magnetospheres. We model energy-dependent light curves (as a function of geometry, e.g., pulsar inclination and observer angle) in the curvature radiation domain using a full emission code. We will discuss our refined calculation of the curvature radius of the particle trajectory and the effect thereof on the expected light curve shapes, as well as the origin of the light curve peaks in the magnetosphere. Our modelling should aid in differentiating between different emission mechanisms, as well as constraining the emission geometry by comparing our predictions to multi-wavelength data.}

\FullConference{5th Annual Conference on High Energy Astrophysics in Southern Africa\\
		4-6 October, 2017\\
		University of the Witwatersrand (Wits), South Africa}

\begin{document}

\section{Introduction}\label{intro}

Pulsar light curves exhibit structure that evolves with photon energy ($E_{\gamma}$). This is a manifestation of the various relativistic particle populations that emit radiation components, as well as the local $B$-field geometry and $E$-field spatial distribution. In addition, Special Relativistic effects modify the emission beam, given the fact that the co-rotation speeds may reach close to the speed of light $c$ in the outer magnetosphere.

Data from ground-based Cherenkov telescopes such as \textit{MAGIC}, \textit{VERITAS}, and \textit{H.E.S.S.-II} that detected pulsed emission from the Crab and Vela pulsars in the very-high-energy (VHE) regime ($>100$~GeV) also exhibit such light curve evolution. \textit{MAGIC} recently detected pulsations from the Crab pulsar at energies up to 1~TeV~\cite{Ansoldi16}, and \textit{H.E.S.S.-II} detected pulsed emission from the Vela pulsar above 100~GeV, making this only the second pulsar to be detected at these high energies \cite{deNaurois15}. Notably, as $E_{\gamma}$ is increased, the main peaks of Crab and Vela seem to remain at the same normalised phase, the intensity ratio of the first to second peak decreases, and the peak widths decrease~\cite{Aleksic12}. Adding data from all $E_{\gamma}$ bands yields an emission spectrum spanning some~20 orders of magnitude~\cite{Harding02,Abdo10C,Abdo10V,Buhler14,Mignani17}.

By constructing detailed physical models, one may hope to disentangle the underlying electrodynamics and acceleration processes occurring in the magnetosphere (see, e.g., the reviews of~\cite{Harding16,Venter17,Venter18}. In this paper we discuss a steady-state emission model~\cite{Harding15} that predicts $E_{\gamma}$-dependent light curves and spectra that result from primary particles emitting curvature radiation (CR). Section~\ref{sec:newrho} briefly summarises our model and the refinement of the calculation of the curvature radius ($\rho_{\rm c}$) of the particle trajectory. In Section~\ref{sec:results} we present our improvement of $\rho_{\rm c}$, sample light curves, and the behaviour of the light curve peaks as a function of $\rho_{\rm c}$, as applied to the Vela pulsar\footnote{The inferred dipolar surface magnetic field for this pulsar is $B_{0}=3.4\times10^{12}$~G; we used a fiducial value of $B_{0}=8\times10^{12}$~G. Uncertainties in $B_0$ will mostly impact the synchrotron emission and not the CR, which is the topic of this paper.}. Conclusions follow in Section~\ref{sec:concl}.

\section{A refined calculation of the curvature radius}\label{sec:newrho}

We use a full emission code that assumes a 3D force-free $B$-field structure and constant $E$-field~\cite{Harding15}. The force-free solution formally assumes an infinite plasma conductivity, so that the $E$-field is fully screened and serves as a good approximation to the geometry of field lines implied by the dissipative models that require a high conductivity in order to match observed $\gamma$-ray light curves~\cite{Kalapotharakos12,Li12,Kalapotharakos14}. 

The primary particles (leptons) are injected at the stellar surface with a low initial speed and are accelerated by a constant $E$-field in a slot gap scenario near the last open field lines. The gap reaches beyond the light cylinder radius $R_{\rm LC}=c/\Omega$ (where the corotation speed equals $c$ with $\Omega$ the angular speed) up to $r=2R_{\rm LC}$. The accelerated primaries radiate CR and some of these $\gamma$-ray photons are converted into pairs causing a pair cascade. This is modelled by injecting a pair spectrum at the stellar surface over the full open volume, without any further acceleration. This pair spectrum is calculated by an independent steady-state pair cascade code using an offset-polar-cap $B$-field that approximates the effect of sweepback of $B$-field lines near the light cylinder~\cite{Harding11}. The pair multiplicity (number of pairs spawned by each primary particle) is kept as a free parameter to allow for the fact that time-dependent pair cascades may yield much larger values for this quantity~\cite{Timokhin15} than steady-state simulations~\cite{Daugherty82}. 

As a first approach we refined the first-order calculation of $\rho_{\rm c}$ along the particle trajectory, assuming that all particles follow the same trajectory, independent of their energy. We assume that the $B$-field is strong enough to constrain the movement of the electrons so they will move parallel to the $B$-field line. Thus, there will be no perpendicular motion in the co-rotating frame since the perpendicular particle energy is instantly expended. Next we take into account the perpendicular $\overrightarrow{E}\times\overrightarrow{B}$ drift (in the lab frame) assuming that the parallel and perpendicular motions of the electrons are independent of the relativistic particle's energy.

To calculate the electron's trajectory as well as its $\rho_{\rm c}$ we used a fixed small step length along the $B$-field line. The first derivative (direction, e.g., bottom panels of Figure~\ref{fig:posdir}) is equivalent to the normalised $B$-field components as a function of the cumulative arclength $s$. First we step along a particular field line. Second, we smooth the directions using $s$ as the independent variable. Third, we match the unsmoothed and smoothed directions of the particle trajectory at particular $s$ values to get rid of unwanted ``tails'' at low and high altitudes, introduced by the use of a kernel density estimator (KDE) smoothing procedure. Fourth, we use a second-order method involving a Lagrange polynomial to obtain the second-order derivatives of the directions along the trajectory as function of $s$. Lastly, we match $\rho_{\rm c}$ calculated using smoothed and unsmoothed directions to get rid of ``tails'' at low and high altitude. We then interpolate $\rho_{\rm c}$  in our particle transport calculations to accommodate the variable step length approach. 

\section{Results}\label{sec:results}

\begin{figure}
\centering
\includegraphics[width=0.8\textwidth]{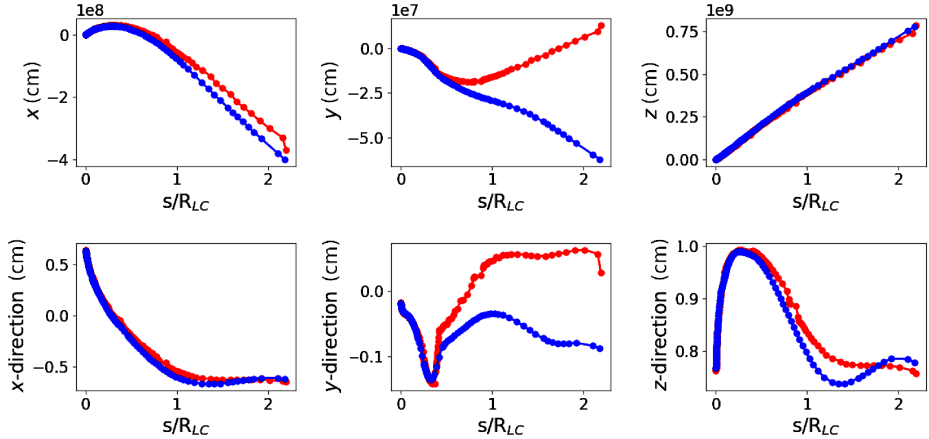}
\caption{The positions (\textit{x}, \textit{y}, and \textit{z} in cm) and directions (first-derivatives; \textit{x}-, \textit{y}-, and \textit{z}-direction in cm) along a specified $B$-field line. Both the previously (red solid-dotted curve) and the more refined approach (blue solid-dotted curve) are indicated, where each dot represents a step along the curved particle trajectory.}\label{fig:posdir}
\end{figure}

\begin{figure}
\centering
\includegraphics[width=.6\textwidth]{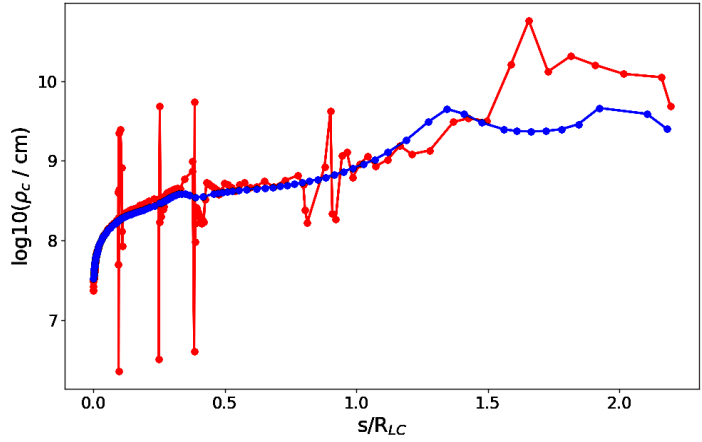}
\caption{The $\log_{10}$ of the previously used $\rho_{\rm c}$ (red solid-dotted curve) and the more refined $\rho_{\rm c}$ (after interpolation; blue solid-dotted curve) along the same $B$-field line as in Figure~\ref{fig:posdir}, where each dot represents a step along the curved particle trajectory.}
\label{fig:rhocomp}
\end{figure}

\begin{figure}
\centering
\includegraphics[width=.9\textwidth]{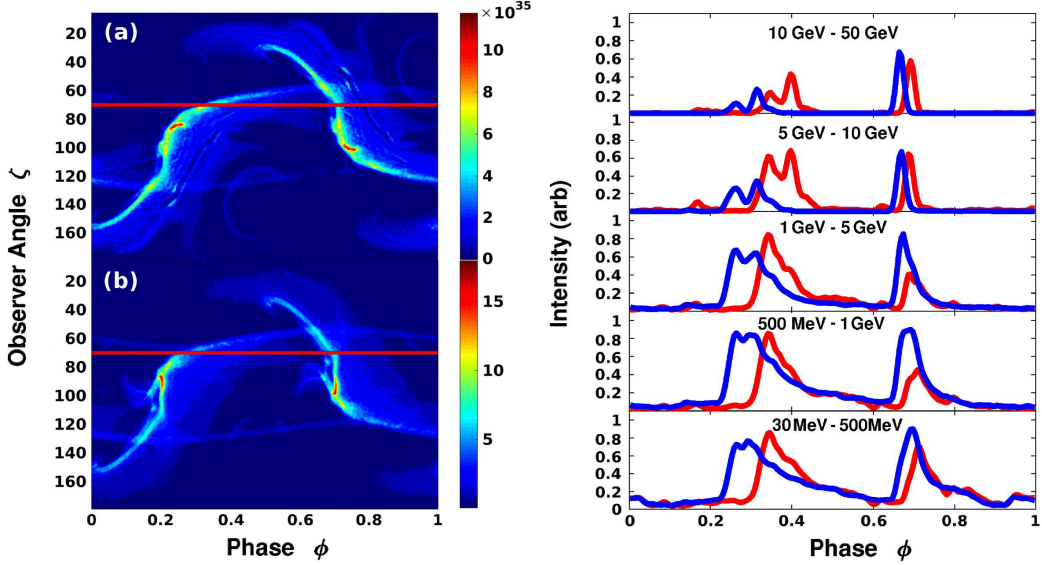}
\caption{Example phase plots (left panels) and light curves (right panels) for $\alpha=45^\circ$, $\zeta=70^\circ$, and $30$~MeV$<E_{\gamma}<50$~GeV. Panel (a) and (b) represent the phase plots for the previously used $\rho_{\rm c}$ and more refined $\rho_{\rm c}$ respectively. Both phase plots have the same relative flux intensity ($\times10^{35}$). On the right is the $E_{\gamma}$-dependent light curves of both the previously used $\rho_{\rm c}$ (red solid line) and more refined $\rho_{\rm c}$ (blue solid line) with $E_{\gamma}$ increasing from bottom to top as indicated by the legend.}\label{fig:exPPLCs}
\end{figure}

In Figure~\ref{fig:posdir} and Figure~\ref{fig:rhocomp} we compare the positions, directions (i.e., first derivatives), and $\rho_{\rm c}$ using the first-order and second-order derivative methods, along a single $B$-field line as a function of $s/R_{\rm LC}$. The deviations of the positions and directions imply that the particle's trajectory follows the $B$-field line more closely when using a second-order method (for variable, large steps, one may veer off a particle trajectory). We thus effectively separated the trajectory and transport calculation. We use a fixed small step length to obtain the particle trajectory and interpolate correctly, and later on we use larger variable step length to perform the transport calculations to save computational time. Figure~\ref{fig:exPPLCs} serves as an example of the effect of the two methods on the phase plots and $E_{\gamma}$-dependent light curves for the Vela pulsar, for an inclination angle $\alpha=45^\circ$, observer angle $\zeta=70^\circ$, and $30~{\rm MeV}<E_{\gamma}<50~{\rm GeV}$. The light curve morphology changes as $E_{\gamma}$ increases. The first peak's relative intensity decreases with respect to that of the second peak, and the second peak becomes narrower with $E_{\gamma}$. The second peak's position remains roughly constant with $E_{\gamma}$. This behaviour is qualitatively similar to that observed by \textit{MAGIC} \cite{Aleksic12} for the Crab pulsar, and by \textit{Fermi} LAT and \textit{H.E.S.S.-II} for Vela \cite{deNaurois15,Abdo10V}. Between the two calculations there is a lag visible between the $\gamma$-ray peaks. 

\begin{figure}[t]
\centering
\includegraphics[width=0.8\textwidth]{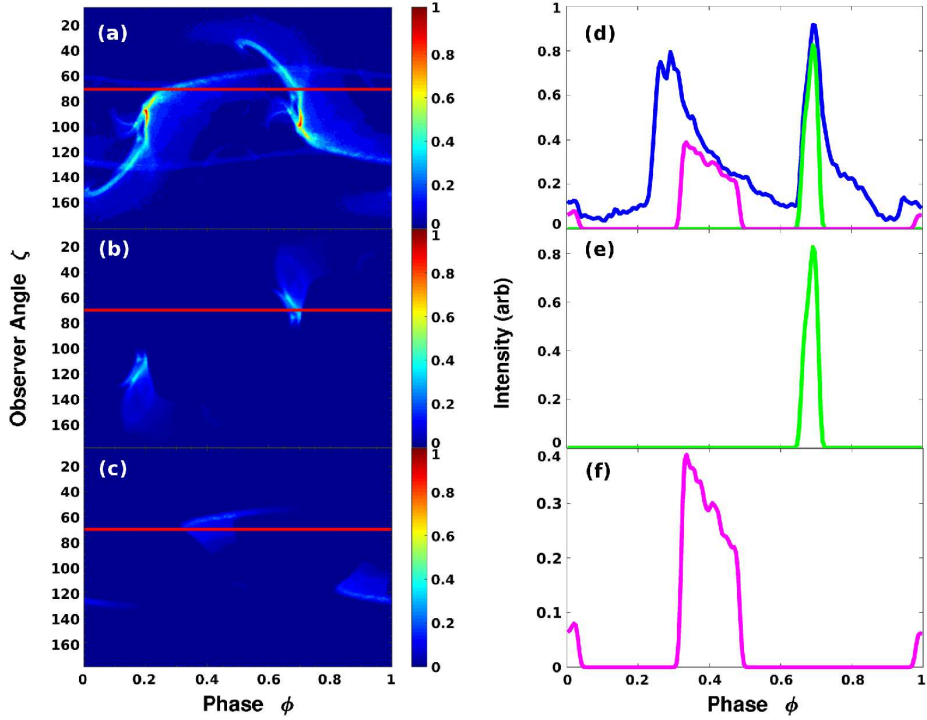}
\caption{Azimuthally-dependent phase plots (left panels) and light curves (right panels) for $\alpha=45^\circ$, $\zeta=70^\circ$, and $30$~MeV$<E_{\gamma}<50$~GeV. Panel (a) and (d) represent the phase plot and its associated light curve for the full PC with $0^\circ<\phi_{\rm PC}<360^\circ$ (blue solid curve). Panel (b) and (e) are for $113^\circ<\phi_{\rm PC}<157^\circ$ (green solid curve) and indicate the origin of peak 2. Panel (c) and (f) are for $315^\circ<\phi_{\rm PC}<360^\circ$ (magenta solid curve) which shows where most of peak 1 originates. Thus, peak 1 originates mostly at $315^\circ<\phi_{\rm PC}<360^\circ$ and peak 2 mostly at $113^\circ<\phi_{\rm PC}<157^\circ$. The phase plots and light curves for the peaks are scaled (i.e., normalised) with the maximum relative flux of the phase plot and light curve for the full PC (panels (a) and (d)), indicating a decrease in relative flux for the peaks.}
\label{fig:delphi}
\end{figure}

\begin{figure}
\centering
\includegraphics[width=0.5\textwidth]{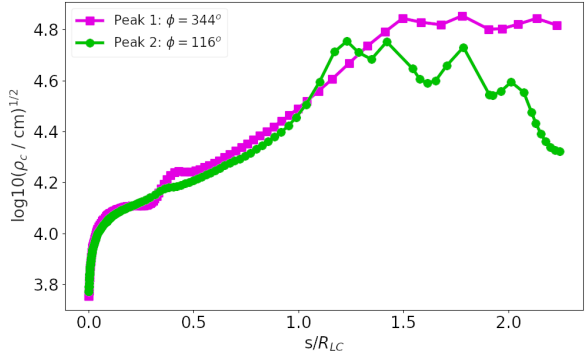}
\caption{Plot of $\log_{\rm 10}(\rho_{\rm c})^{1/2}$ for typical particle orbits associated with peak 1 (magenta curves) and peak 2 (green curves).}
\label{fig:rhohalf}
\end{figure}

In order to pin down the location where the emission originate, we need to roughly transform $\rho_{\rm c}(\overrightarrow{r})$ to $\rho_{\rm c}(\phi_{\rm PC})$. The azimuthal angle $\phi_{\rm PC}$ is the angle measured on the stellar surface about the magnetic axis, which is different from the rotation phase $\phi$. We show caustics and light curves in Figure~\ref{fig:delphi} for $\alpha=45^\circ$, $\zeta=70^\circ$, and $30~{\rm MeV}<E_\gamma<50~{\rm GeV}$ originating from different sections of the polar cap (PC). These include $\phi_{\rm PC}$ ranges $0^\circ<\phi_{\rm PC}<360^\circ$ (the full PC), $113^\circ<\phi_{\rm PC}<157^\circ$, and $315^\circ<\phi_{\rm PC}<360^\circ$. Since each $B$-field line has a unique footpoint (associated with a specific $\phi_{\rm PC}$) on the PC at the stellar surface, the $E_{\gamma}$-dependent light curve morphology is also azimuthally dependent. From these light curves we find that the emission from the first peak originates mostly\footnote{In future we will refine our choices of the $\phi_{\rm PC}$ intervals, so as to obtain closer matches to peak 1 and peak 2 for different values of $\alpha$ and $\zeta$ (see Figure~\ref{fig:delphi}).} at $315^\circ<\phi_{\rm PC}<360^\circ$ and the second peak mostly at $113^\circ<\phi_{\rm PC}<157^\circ$. We speculate that the vanishing of peak 1 stems from the fact that the two peaks originate in regions of the magnetosphere that contains $B$-field lines characterised by slightly different curvature radii $\rho_{\rm c}$ (see Figure~\ref{fig:rhohalf}). This must be the case since we have assumed a constant accelerating $E_\parallel$-field in this paper. 

\begin{figure}
\centering
\includegraphics[width=0.5\textwidth]{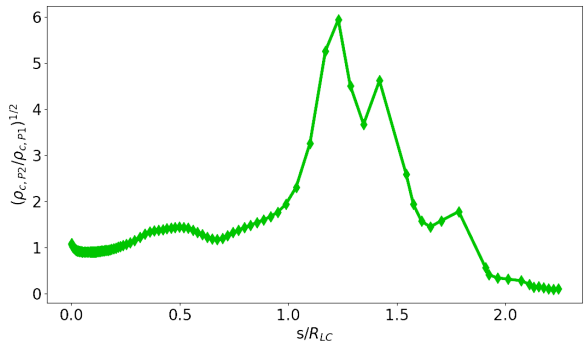}
\caption{The ratio between $\rho_{\rm c}^{1/2}$ for peak 1 and peak 2 remains roughly constant inside the light cylinder and then abruptly increases at and beyond the light cylinder (i.e., in the current sheet).}
\label{fig:ratio}
\end{figure}

In the CR reaction (CRR) limit, where the particle acceleration rate equals the CR loss rate, we find
\begin{equation}
\gamma_{\rm RR} = \left(\frac{3E_{\parallel}\rho_{c}^2}{2e\beta_{r}^{3}}\right)^{1/4}.
\end{equation}
This implies
\begin{equation}
E_{\gamma, {\rm cutoff}} \sim 4E_{\parallel,4}^{3/4} \rho_{c,8}^{1/2} \quad {\rm GeV},
\end{equation}
which scales with $\rho_{\rm c}^{1/2}$ (see \cite{VenterDeJager2010} for definition of symbols){\footnote{For the Vela pulsar, assuming a constant inverse acceleration length $R_{\rm acc}=0.2$~cm$^{-1}$ and thus accelerating $E_\parallel=m_{\rm e}c^2R_{\rm acc}/e=341$~G, we find $\gamma_{\rm RR}\approx{1.02\times10^7\rho_{c,8}^{1/2}}$, where $\rho_{c,8}=\rho_c/(10^8$~cm).}}. In the CR regime we expect that the CRR limit must be reached. Even if this limit is not attained, each peak's spectral cutoff $E_{\gamma, {\rm cutoff}}$ should still depend on the local range of $\rho_{\rm c}$ where this emission originates. Peak 2 with the larger $\rho_{\rm c}$ should have a larger $E_{\gamma, {\rm cutoff}}$ (see Figure~\ref{fig:rhohalf} and~\ref{fig:ratio}; Barnard et al., in prep.).

\section{Conclusions}\label{sec:concl}

Modelling of $E_{\gamma}$-dependent pulsar light curves as well as their spectra is vital to disentangle the effects of acceleration, emission, beaming, and $B$-field geometry. We used a 3D emission model assuming CR from primary particles in an SG reaching $2R_{\rm LC}$ to study the evolution of the predicted light curves in different $E_{\gamma}$ bands. We find that emission from beyond $R_{\rm LC}$ (in the current sheet, e.g., \cite{Bai10}) constitutes an important contribution to the light curve structure. We also observe that the predicted ratio of the first to second peak intensity decreases\footnote{This behaviour was also seen for the retarded vacuum dipole field (Rudak, private communication).}. The second peak becomes narrower with increasing $E_{\gamma}$, and its position in phase remains steady with $E_{\gamma}$, similar to what has been observed at $\gamma$-ray energies for the Crab and Vela pulsars. 

The refinement of $\rho_{\rm c}$ changed the phase plots and light curves slightly. We find that the origin of the light curve peaks are both altitude- and azimuthally-dependent. The $\rho_{\rm c}$ is greater for peak 2 than peak 1, leading to a greater $E_{\gamma, {\rm cutoff}}$ for peak 2. This may explain phenomena seen by \textit{Fermi} and \textit{H.E.S.S.-II}.

It is not clear what the emission mechanism for high-energy light curves is. The standard models assumed this to be CR (e.g., \cite{Daugherty96,Romani96}), while newer models focus on synchrotron radiation in the current sheet \cite{Petri12,Philippov15,Cerutti16,Philippov18}. Continued spectral, light curve and now polarisation modelling \cite{Cerutti16,Harding17}, confronted by quality measurements, may provide the key to discriminate between different models.

\acknowledgments
This work is based on the research supported wholly / in part by the National Research Foundation of South Africa (NRF; Grant Numbers 87613, 90822, 92860, 93278, and 99072). The Grantholder acknowledges that opinions, findings and conclusions or recommendations expressed in any publication generated by the NRF supported research is that of the author(s), and that the NRF accepts no liability whatsoever in this regard. A.K.H. acknowledges the support from the NASA Astrophysics Theory Program. C.V. and A.K.H. acknowledge support from the \textit{Fermi} Guest Investigator Program.

\end{document}